# Chemical gradients across phase boundaries between martensite and austenite in steel studied by atom probe tomography and simulation

O. Dmitrieva[(1)], D. Ponge[(1)], G. Inden[(1)], J. Millán[(1)], P. Choi[(1)], J. Sietsma[(2)], D. Raabe[(1)]

(1) Max-Planck-Institut für Eisenforschung, Max-Planck-Str. 1, 40237 Düsseldorf, Germany
(2) Delft University of Technology, Faculty 3mE, Dept. MSE, 2628 CD Delft, The Netherlands

o.dmitrieva@mpie.de, d.ponge@mpie.de, d.raabe@mpie.de

## Abstract

Partitioning at phase boundaries of complex steels is important for their properties. We present atom probe tomography results across martensite / austenite interfaces in a precipitation-hardened maraging TRIP steel (12.2 Mn, 1.9 Ni, 0.6 Mo, 1.2 Ti, 0.3 Al; at.%). The system reveals compositional changes at the phase boundaries: Mn and Ni are enriched while Ti, Al, Mo, and Fe are depleted. More specific, we observe up to 27 at.% Mn in a 20 nm layer at the phase boundary. This is explained by the large difference in diffusivity between martensite and austenite. The high diffusivity in martensite leads to a Mn-flux towards the retained austenite. The low diffusivity in the austenite does not allow accommodation of this flux. Consequently, the austenite grows with a Mn-composition given by local equilibrium. The interpretation is based on DICTRA and mixed-mode diffusion calculations (using a finite interface mobility).

## 1. Introduction

Mn is among the most important alloying elements for the design of advanced high strength steels, as it affects the stabilization of the austenite, the stacking fault energy, and the transformation kinetics [1-11]. Besides these global mechanisms which are exploited particularly in designing steels with TRIP and TWIP effects, Mn has very low diffusion rates in the austenite and a high segregation or respectively partitioning tendency at interfaces (TRIP: transformation-induced plasticity; TWIP: twinning-induced plasticity). This context makes Mn (besides the other elements discussed in this paper) a very interesting candidate for an atomic-scale study of compositional changes across austenite/martensite interfaces.

The specific material studied in this work is a precipitation-hardened alloy that we refer to as maraging TRIP steel. It was developed by combining the TRIP mechanism with the maraging effect (maraging: martensite aging) [12,13]. The TRIP effect exploits the deformation-stimulated transformation of metastable retained austenite into martensite and the resulting plasticity required to accommodate the transformation misfit [1-7]. The maraging effect uses the hardening of the heavily strained martensite through the formation of nano-sized intermetallic precipitates during aging heat treatment. The maraging TRIP steels used in this work reveal the surprising property that both strength and total elongation increase upon aging reaching an ultimate tensile strength of nearly 1.3 GPa at an elongation above 20% [12-14].

The studied alloy contains 12.2 at.% Mn, low carbon content (0.05 at. %) and minor additions of Ni, Ti, Al, and Mo. Its microstructure after aging is characterized by the presence of up to 15-20 vol.% austenite, a fine martensite matrix, and dispersed nanoscaled Ni-Al-Mn-enriched zones [12-14]. Besides the increase in strength, a simultaneous increase of ductility was found upon aging. This effect is interpreted in terms of sluggish re-austenitization during aging and the effect of tempering of the as-quenched martensite [14]. Partial re-transformation into austenite (besides the existing retained austenite) by a reconstructive mechanism involving manganese partitioning might be responsible for this process.

In order to elucidate this transformation phenomenon, particularly the role of Mn, we focus in this work on the analysis of nanoscopic elemental diffusion gradients across abutting martensite / retained austenite phase areas. Atom probe tomography (APT) is a characterization technique that provides three-dimensional elemental mapping with nearly atomic resolution and gives information on the topology of interfaces and local chemical gradients [15-27]. We conducted atom probe tomography using an advanced local electrode atom probe device (IMAGO LEAP 3000X HR). Both the two phases (austenite, martensite) and the interfaces between them were chemically analyzed at the atomic scale. Additionally, statistical thermodynamic and kinetic calculations were conducted for the given initial and boundary values using Thermo-Calc [28,29] in conjunction with the kinetic simulation software DICTRA [30-32] and with a mixed-mode kinetic approach that considers finite interface mobility [33,34].

## 2. Experiments

The investigated maraging TRIP steel with a composition of 12.2 at.% Mn, 1.9 at.% Ni, 0.6 at.% Mo, 1.2 at.% Ti, 0.1 at.% Si, 0.3 at.% Al, and 0.05 at.% C was melted and cast in a vacuum induction furnace. Before final age hardening, a solution treatment was performed in Ar atmosphere at 1050 °C for 0.5 h followed by water quenching. This led to a microstructure consisting of martensite and retained austenite. Final aging was conducted for 48 hours at 450 °C. After aging the sample was quenched in water. Details of the alloy preparation were published elsewhere [12-14].

APT samples were prepared by electrochemical polishing and subsequent sharpening using a focused ion beam device. Pulsed-laser APT was performed using a local electrode atom probe (LEAP$^{TM}$ 3000X HR, Imago Scientific Instruments) tomograph at a specimen temperature of 54 K. An ultrafast pulsed laser of ~ 10 ps pulse width and 532 nm wavelength was applied at a frequency of 250 kHz. The laser pulse energy was set to 0.4 nJ. The detection rate (target evaporation rate) amounted to 5 atoms per 1000 pulses. Data analysis was performed using the IVAS$^{®}$ software from Imago Scientific Instruments. The specific APT data set analyzed in this work contains about 70 million ions. We used an evaporation field constant of 26 V/nm for the atomic reconstruction.

Phase fractions and the elemental compositions in thermodynamic equilibrium were calculated using the software Thermo-Calc [28]. The software DICTRA [30-32] and a mixed-mode kinetic approach including finite interface mobility [33,34] were applied to simulate diffusion controlled phase transformations. The simulations were performed using the thermodynamic database TCFE6 [29] and the mobility database MOB2.

## 3. Results

### 3.1. Analysis of the 3D atom probe reconstruction

3.1.1. Manganese distribution

Fig. 1a gives a microstructure overview of the maraging-TRIP steel after quenching and subsequent aging (48 at 450 °C). The upper micrograph is an EBSD image where the cubic martensite is plotted green and the retained austenite red (the retained austenite was already present in the as-quenched state before aging [12-14]). The middle image shows a TEM micrograph with precipitate-containing martensite and precipitate-free austenite. The bottom image shows an APT reproduction which includes both martensitic and austenitic zones. Ni atoms are given in cyan and Mn atoms in blue. The yellow iso-surfaces indicate 18 at.% Mn. Note that the three images reveal the hierarchy of the microstructure but the individual images were not exactly taken at the positions indicated. Fig 1b gives a local overview of the distribution of the Ni and Mn atoms in the center of the APT data set presented in Fig. 1a. For clarity, only a longitudinal section of 20 nm thickness is shown, in which only 7.8 % of all detected Ni (cyan) and 1.5% of all Mn (dark blue) atoms are displayed. The whole analysis volume is about $4\times10^5$ nm$^3$. Figs. 1a,b show three main zones that are separated by inclined plate-like Mn accumulations. Ni-rich nanoprecipitates are dispersed in the left and right hand areas. Besides the Ni atoms, also higher amounts of Al, Mn, and Ti were detected in these clusters, Table 1. In the center part between the Mn enriched plates, no precipitates appear. This observation strongly suggests that this zone corresponds to austenite, whereas the abutting areas containing precipitates are martensitic. Correlative TEM investigations conducted on this alloy in the same aging state (48h, 450°C) support the suggestion that the nano-particles that are enriched in Ni, Al, and Mn formed were in the martensitic microstructure while the retained austenite (total volume fraction about 15-20 vol.%) was precipitate-free [13,14], Fig. 1a. From these observations we conclude that the present volume probed by APT contains an austenitic grain enclosed between two martensitic grains. Quantitative chemical analysis of the interfaces between austenite and martensite was performed using 1D concentration profiles computed over the region of interest (transparent cylindrical units), Fig. 2a. We calculated the Mn content averaged over the 0.5 nm thick cross sections of the cylinders at a profile step size of 0.5 nm. For both interfaces, a strong increase in the Mn content up to 27 at. % was observed, Fig. 2b. Away from the interface, the content of Mn within the austenite amounts to about 12 at. % which is close to the average chemical composition of the alloy. Within the bulk martensite the Mn content amounts to about 10 at. %. Mn depletion in the martensite down to 6 at. % was observed close to the interface.

In order to exclude the contribution of the precipitates from the chemical profile within the martensitic area, we separately measured the 1D concentration profiles within the martensitic matrix after removing the precipitate zones from the analysis volume (described in detail below). The reason for this procedure is that the martensitic area is in itself a two-phase region consisting of martensite and precipitates. Hence, we aim with this method at the separation of the martensite elemental composition and the precipitate elemental composition. These two corrected profiles, containing only the martensite composition, are included in Fig. 2b on the left hand side in the martensitic area marked 'M'. The curves are separated from the profile across the interface and in the austenite ('A').

3.1.2. Distribution of other alloying elements

Fig. 3 shows the concentration profiles for the other alloying elements across one of the martensite/austenite phase boundary zones. The area selected is indicted by 'Mn layer 2' in Fig. 2a. Besides Mn (which is studied here in more detail owing to its relevance for high strength steels) also all other elements reveal a strong partitioning between the two phases. While Mn is enriched by about 2.1 times within the interface boundary layer relative to its average content in the alloy, Ni is accumulated 1.2 times in the same zone. All other elements are depleted in the interface zone, namely, Ti decays by a factor of about 6.9 times relative to the average content, Al by a factor of 6.6, Mo 2.0, and Fe 1.2. Another important observation is the large chemical width of the phase boundary zone: The enrichment zone associated with the austenite/martensite interface extends over a length of about 20 nm normal to the boundary segment studied.

3.1.3. Chemical analysis of the nanoparticles and of the alloy matrix

The nanoparticles detected in the martensite were analyzed using a cluster search algorithm implemented in the IVAS® software. For cluster identification, the following parameters as identified by the optimization procedure performed within the cluster search algorithm were used: $d_{max}$ = 0.6 nm (maximal distance between the solute atoms belonging to a cluster), $N_{min}$ = 50 (minimal number of solute atoms in the cluster), L = 0.57 nm (envelope distance: all non-solute ions within a distance L of solute ions are included in the cluster), $d_e$= 0.55 nm (erosion distance: all clustered non-solute ions within a spacing $d_e$ of any ion outside of its assigned cluster are removed from the particle). The cluster search was conducted for the distribution of the Ni atoms that are enriched in the particles. The chemical composition of the clusters is summarized in Table 1. The calculation of the enrichment factors that were determined as a relation between the content within the particles relative to the total content of the same elements in the entire alloy reveals a strong precipitation character of Ni and Al atoms within the clusters. Enrichment in Mn and Ti was also detected in the particles. For estimating the chemical composition of the surrounding matrix (without the precipitates), the detected clusters were removed from the overall reconstruction, and the composition of the residual matrix volume was calculated again. For this purpose, different cluster search parameters were used ($d_{max}$ = 1.0 nm, $N_{min}$ = 50, L = 0.97 nm, $d_e$ = 0.95 nm) which allow to include more material in the clusters and to ensure that after exclusion of the clusters no residual material remains. The composition of the matrix without the particles is presented in Table 1.

3.1.4. Observation of compositional changes within the martensite/austenite interface region

By computing the iso-concentration surfaces for all solute elements from the experimental data we detected changes in composition of Mo, Ti, and Si in the martensite/austenite interface region. Fig. 4a shows iso-surfaces for Mo, Ti, and Si concentrations of 5 at.%, 8 at.%, and 3 at.%, respectively. The position of this region overlaps with the position of the iso-concentration surface plotted at 18 at.% Mn and, more specific, corresponds to the range of the highest Mn gradient. A similar region with nearly the same content and element distribution was also observed at the other martensite/austenite interface (not shown in Fig 4a). The average chemical composition within the region is summarized in the table in Fig. 4b. The enrichment factors reveal strong compositional increase of Ti, Mo, and Si and strong depletion of Al. The relative concentrations of Fe, Ti, and Mo within that region as 75:17:8 suggest the formation of a Laves phase, which should be formed according to Thermo-Calc at a composition of Fe 67 at.%, Ti 23 at.%, and Mo 10 at.%.

**3.2. Thermodynamic calculations**

3.2.1. Prediction of the phase equilibrium composition

Using Thermo-Calc the equilibrium compositions of stable phases at 450°C were calculated taking into account the total nominal composition of the alloy and all possible competing phases available in the database [28,29]. The results of the Gibbs energy minimization technique predicts four phases in thermodynamic equilibrium, namely, BCC (ferrite/martensite), FCC (austenite), TiC, and a Laves phase. The calculated molar fractions for each phase and their chemical compositions are listed in Table 2. It is important to point out that such calculation does not predict the presence of the nano-sized particles due to the limited availability of thermodynamic data related to various other possible intermetallic phases in complex maraging steels. At a temperature of 450°C, a Mn-content of about 26.7 and 3.3 at.% are expected in the retained austenite and in the ferrite (martensite), respectively.

3.2.2. Diffusion simulations using DICTRA

For the simulation of the kinetic behavior in the vicinity of the martensite/austenite interface the linear cell geometry is appropriate [30]. The kinetic effects to be studied are confined to very small spatial ranges. Within this scale the interfaces are planar in shape and their movement is vertical to the plane. The size of the cell was chosen as 20μm, see Fig. 5a. The cell was divided into two regions, one corresponding to ferrite, the other one to austenite. The space in each region is discretized as a linear grid. The distribution of the grid points is chosen with a high density close to the interface. The grid is defined in terms of geometric series. The compositions of ferrite and austenite were taken according to the values determined via the APT characterization for the austenitic and martensitic matrices, respectively. In the martensitic matrix we detected a slight Mn depletion down to 10.3 at. %. This can be attributed to the formation of nano-precipitates. Within martensite a high amount of lattice defects, particularly dislocations, enhance the atomic diffusion in this phase. Our previous TEM-based studies on this material [12,13] revealed that most of the nano-precipitates were indeed associated with dislocations which supports the assumption that the pipe-mechanism may strongly assist diffusion within the martensite. In order to take into account the variation of composition in the ferrite due to the precipitation of particles, an average composition was used for the ferrite phase (see Fig. 5b). Martensite is not included as a separate phase in the thermodynamic and kinetic databases as the thermodynamic properties of martensite are very much the same as those of ferrite. Therefore, in the thermodynamic calculations, martensite is represented as ferrite. The kinetic parameters, however, may deviate between the two phases owing to the defect structure and distortion of the martensite. Up to now, no detailed information is available on the effect of these conditions on possible changes in the kinetic parameters between ferrite and martensite.

The size of the cell is fixed during the simulation (20μm), whereas the interface between the two regions is mobile. The conditions at the moving interface are determined by the local equilibrium assumption, i.e. the chemical potentials of all diffusing elements assume the same value in ferrite and austenite. The value of the potentials is controlled by the mass balance condition. Diffusion of Mn, Mo, Ni, Si, and Ti atoms was considered in the calculation. The simulation was performed for the aging temperature 450°C and stopped at time step 180000 sec (50h).

The composition profile of Mn between ferrite and austenite after an annealing time of 50h at 450°C is presented in Fig. 6. The interface has moved towards the ferrite side leaving behind an austenite layer with drastically changed composition. This result is in qualitative agreement with the experimental data presented in Fig. 2b. However, the width of the predicted Mn-rich interfacial layer is too small and, correspondingly, the extent of the Mn depletion zone in ferrite is relatively small, too. This discrepancy indicates that the mobility of Mn in the martensitic matrix must be higher than it is in ferrite. Therefore, the simulations were repeated with increased mobilities of the elements in the ferrite. The results of the simulations with a factor 12 and 45, respectively, for the enhanced mobilities in martensite are shown in Fig. 7. The result in Fig. 7b (45 times enhanced Mn mobility in martensite) is in excellent agreement with the experimental results in Fig. 2b. This applies for the depletion profile of Mn in the martensite and also for the Mn-enriched interface zone.

In view of this good agreement the profiles of the other elements should also be analyzed. Fig. 8a shows the composition profiles of some of the elements for the case of the mobility factor 45. Fig. 8b presents the predicted enrichment or depletion of the other elements, respectively, in the same way as for the experimental results. The partitioning tendencies of the elements are the same as observed in the experiment (compare Figs. 3a,b and Figs. 8a,b). The predicted enrichment of Mn and Ni and the depletion of Mo within the interfacial austenite layer are in good quantitative agreement with the experiments. For Ti and Si the decrease is less pronounced in the simulation than in the experiment.

We estimated the mean diffusion paths of Mn atoms in both phases using the diffusion coefficients obtained for 450°C using DICTRA (Mob2 database). The diffusion constant of Mn atoms in a BCC iron matrix (ferrite) was $D_{BCC} = 1.75 \times 10^{-22}$ m$^2$/s and in the FCC iron matrix $D_{FCC} = 5.86 \times 10^{-24}$ m$^2$/s. The mean diffusion path $\lambda$ of Mn atoms for an aging time of $t = 48$h was calculated using the volume diffusion equation for cubic metals: $\lambda = (6tD)^{\frac{1}{2}}$. The mean diffusion path of Mn atoms in the BCC lattice was about 13 nm and in the FCC lattice only about 2.5 nm. Thus, the diffusion length of Mn in BCC is significantly larger than in the FCC lattice. When correcting the mobility of the atoms by a factor of 45, as explained above, the diffusion constant in ferrite (which can be then treated as martensite) is $7.56 \times 10^{-21}$ m$^2$/sec. For this case, the mean diffusion path of Mn in BCC increases from 13 nm to about 90 nm.

## 4. Discussion

### 4.1. Phase boundary motion with infinite interface mobility in the DICTRA approach

The global equilibrium calculated with Thermo-Calc (see Table 2) predicts a high amount of Mn in the retained austenite (27.6 at.%) and a low value (3 at.%) in ferrite. Hence, during aging a redistribution of Mn is expected. However, the global equilibrium only indicates the long term trends. The actual situation at the phase boundary is controlled by a local equilibrium.

It is not possible to visualize graphically equilibria in multicomponent systems. Therefore, the following discussion will be done considering only three components, Fe, Mn, Ni. Fig. 9 shows the ternary phase diagram at 450°C. The initial compositions of martensite (filled square) and austenite (filled circle) are within the two-phase region α+γ. The global equilibrium tie-line is shown by a dotted line. The slope of the tie-lines indicates that at the austenite phase boundary the level of both Mn and Ni must be higher than in the matrix. Conversely, the level of Mn and Ni at the martensite boundary must be lower than in the matrix. The range of possible local equilibrium tie-lines is thus confined to those originating from Ni-concentrations in martensite below the value in the α-matrix, i.e. $x_{Ni}^{\alpha/\gamma} < x_{Ni}^{\alpha}$. This composition range is marked in Fig. 9 by a bold phase boundary. The operating local equilibrium tie-line is defined by the fluxes of Ni and Mn. The interface displacement caused by these fluxes must be the same for every diffusing element. The resulting operating local tie-line is indicated in Fig. 9, showing the difference to that of global equilibrium.

Due to the low diffusivity in austenite, the fluxes lead to an interface displacement towards martensite. The layer of increased Mn is the result of the partitioning imposed by the local equilibrium tie-line during the formation of austenite. An epitaxial formation of this aging-induced austenite at the phase boundary of the existing austenite is likely.

The overall agreement between experiment and simulation is very good. There is a slight difference in the Mn-composition at the martensite boundary though. The experiments yield a value of about 5-6 at.% Mn, while the simulation gives a value of 3 at.%. There are two possibilities to explain this difference. At first, it can be an effect of the resolution of the experiment. The transition from the low concentration at the martensite boundary to the very high concentration at the austenite interface occurs sharply. It is, therefore, plausible that close to this abrupt transition the Mn signal is slightly contaminated by the elevated Mn concentration, leading to a slightly increased composition close to the boundary. Another possibility could be a finite mobility of the interface. The local equilibrium approach implies that the interface can move freely. A finite mobility of the interface (see details in next section) leads to a slower interface velocity than that given by the local equilibrium. Consequently, the boundary condition has to be adapted such that the mass balance is in accordance with the velocity. This effect, however, should become more relevant at rather high interface velocities. In the present case of low temperature the interface velocity is small, see Fig.10. This velocity range is more than six orders of magnitude smaller than usual velocities occurring during the transition between austenite and ferrite at temperatures of about 700°C or higher. We, hence, anticipate that the finite interface mobility effect might play a second-order role in the current case as discussed in more detail in the next section.

**4.2 Phase boundary motion with finite interface mobility in the mixed-mode approach**

The simulation of a partitioning phase transformation with DICTRA as outlined above is based on the assumption that Mn diffusion is controlling the transformation kinetics. In a more generalized mixed-mode approach [33,34] the motion of the interface during the transformation is defined by its velocity v, given by

(1) $\quad v = M \Delta G$

with ΔG the free-energy difference between the phases, acting as the driving force for transformation, and M the interface mobility. In the purely diffusion-controlled transformation, such as discussed above, M is assumed to be infinite, which means that the interface instantaneously reacts to any deviation of the local concentration from equilibrium, thus restoring the local equilibrium. If M is finite, though, a certain balance is established between diffusion (in the case of Mn in α increasing the interface concentration, which increases the driving force) and interface motion (decreasing the Mn-concentration at the interface, which decreases the driving force). For given values of M and the diffusivity D, the resulting value of the interface concentration, and thus of the driving force and the velocity, can be simulated for binary Fe-Mn on the basis of two assumptions [33]. The first one is that the driving force is proportional to the deviation from equilibrium

(2) $\quad \Delta G = \chi \left( x_{Mn}^{\alpha\gamma} - x_{Mn}^{\alpha} \right)$

with the indices 'αγ' denoting the equilibrium concentration in the α-phase (martensite) in equilibrium with γ (austenite), and $x_{Mn}^\alpha$ the Mn-concentration in martensite at the interface. In the case of partitioning Mn the proportionality factor $\chi$ is negative. The second assumption is that the diffusion in the parent martensitic α-phase leads to a concentration profile that can be described by an exponential function

$$(3) \quad x_{Mn} = x_0 + \left(x_{Mn}^\alpha - x_0\right)\exp\left(-\frac{z}{z_0}\right)$$

with the spatial co-ordinate $z = 0$ at the interface. The width parameter $z_0$ follows from the values of $M$ and $D$ and the equilibrium and overall ($x_0$) concentrations [33]. Diffusion of Mn in the austenite is so slow that it can be neglected.

The experimental profile in Fig. 2b shows a Mn concentration in the α-phase at the interface of 5-6 at.%, which is slightly larger than the equilibrium value of 3.3 at.%. This would imply a deviation from local equilibrium. Using the value of the interface mobility $M$ as an adaptable parameter and the same enhanced Mn diffusivity (factor 45) as used in the DICTRA simulations above, the Mn-profile in the martensitic phase can be adequately reproduced, Fig. 11. The final profile ($t = 180000$ s) is given, but also an intermediate stage, after 28000 s. It is revealed that the deviation from equilibrium is larger in the earlier stages of the transformation. The calculations were conducted for a value of $\chi = -12.8$ kJ/mol, determined with ThermoCalc, and a mobility of $M = 2\times10^{-21}$ m$^4$/Js at $T = 450°$C. The simulated results reveal an excellent agreement with the experiments, Fig. 11. The mobility value, however, is distinctly smaller than the mobility data for the standard γ→α transformation around $T = 800°$C, when extrapolating with the commonly used activation energy for the mobility of 140 kJ/mol.

The mixed-mode approach uses the interface mobility as an adaptable parameter. This approach is particularly useful in cases where the transition is not fully controlled by diffusion. In such cases the local chemical equilibrium condition cannot be fulfilled. Instead, a difference in chemical potentials exists at the interface which provides the Gibbs energy required for the motion of the interface. If the interface mobility is known and does act as a limiting kinetic factor, it may play an essential role in the formation of the overall microstructure and hence should be included in corresponding predictions.

**4.3 Comparison and conclusions from the two simulation methods**

The Mn distributions predicted by the calculations revealed diffusion of Mn from the ferritic phase towards the austenitic matrix and the accumulation of Mn at the interface between these two regions. The composition profiles obtained experimentally agree with the simulations provided that the mobility of all alloying elements in martensite are increased compared to ferrite (by a factor of 45, Fig. 7b). This applies to both types of simulation approaches, DICTRA [30-32] and mixed-mode [33,34]. Such an enhanced diffusion in martensite can be attributed to a high defect concentration (e.g. misfit dislocations introduced through the transformation) in martensite. Pipe diffusion might hence be one reason for this enhanced diffusion [12-14].

So far it is not clear whether the higher mobility is valid for martensite in general or if this holds only in the neighborhood of the phase boundary which may act as a source of vacancies and provides high local dislocation densities in its vicinity [35]. More experimental information is needed to elucidate this point.

# 5. Conclusions

We studied compositional variation phenomena on martensite/austenite interfaces in a maraging TRIP steel. We placed particular attention on the partitioning of Mn at these interfaces using 3D atom probe analysis in conjunction with ThermoCalc, DICTRA, and mixed-mode simulations (where the latter one also includes the heterophase interface mobility). The local boundary condition at the interface leads to the diffusion of Mn in martensite towards austenite. The chemical gradients of Mn predicted by DICTRA at the phase boundary revealed a good quantitative correlation to the experimental findings. The diffusion behavior of other alloying elements such as Ni, Ti, and Mo could also be reproduced in the dynamic simulation.

The partitioning at the martensite/austenite interface leads to the formation and growth of a new austenite layer on the existing retained austenite with drastically changed composition compared to the bulk. It is to be expected that such a layer will have an effect on the mechanical properties. In the present case, this layer is likely to be one of the microstructural changes during aging that might be responsible for the unexpected increase in ductility after the annealing treatment [12]. The other contribution for increasing the ductility stems from the tempering of the martensitic matrix during annealing and was reported elsewhere [36].

By using the advanced APT technique we gained deep insights into the chemical nature and dynamics of the martensite/austenite phase boundary during aging. The theory-assisted 3D chemical analysis at the nanoscale provides significant enhancement of our understanding of partitioning affects and their relationship to phase transformation kinetics in multi-phase steels.

## References

[1] Patel JR, Cohen M. Acta Metall 1953;1:531.
[2] Bhadeshia HKDH, Edmonds DV. Metall Trans 1979;10A:895.
[3] Takahashi M, Bhadeshia HKDH. Mater Trans JIM 1991;32:689.
[4] Jacques PJ, Girault E, Catlin T, Geerlofs N, Kop T, van der Zwaag S, Delannay F. Mater Sci Eng 1999;A273-275:475.
[5] De Meyer M, Vanderschueren D, De Cooman BC. ISIJ Int 1999;39:813.
[6] Traint S, Pichler A, Hauzenberger K, Stiaszny P, Werner E. Steel Res Int 2002;73:259.
[7] Zaefferer S, Ohlert J, Bleck W. Acta Mater 2004;52:2765.
[8] Brüx U, Frommeyer G, Grässel O, Meyer LW, Weise A. Steel Res 2002;73:294.
[9] Tomota Y, Strum M, Morris JW. Metall Mater Trans 1986;17A:537.
[10] Song R, Ponge D, Raabe D, Kaspar R. Acta Mater 2004;53:845.
[11] Song R, Ponge D, Raabe D. Acta Mater 2005;53:4881.
[12] Raabe D, Ponge D, Dmitrieva O, Sander B. Scripta Mater 2009;60:1141.
[13] Raabe D, Ponge D, Dmitrieva O, Sander B. Adv Eng Mater 2009;11/7:547.
[14] Ponge D, Millán J, Dmitrieva O, Sander B, Kostka A, Raabe D, Proc 2nd Inter Symp Steel Sci (ISSS 2009), Kyoto, Japan: The Iron and Steel Institute of Japan; 2009, p. 121.
[15] Cerezo A, Godfrey TJ, Smith GDW. Rev Sci Instrum 1988;59:862.
[16] Blavette D, Deconihout B, Bostel A, Sarrau JM, Bouet M, Menand A. Rev Sci Instrum 1993;64:2911.
[17] Miller MK, Cerezo A, Hetherington MG, Smith GDW. Atom Probe field ion microscopy. Oxford UK: Oxford University Press; 1996.
[18] Thuvander M, Miller MK, Stiller K. Mater Sci Eng A 1999:270:38.
[19] Miller MK. Atom probe tomography analysis at the atomic scale. New York: Kluwer Academic/Plenum Publ; 2000.
[20] Kelly TF, Miller MK. Rev Sci Instrum 2007;78:031101.
[21] Seidman D. Annu Rev Mater Sci 2007;37:127.
[22] Miller MK, Forbes RG. Mater Character 2009;60:461.
[23] Marquis EA, Miller MK, Blavette D, Ringer SP, Sudbrack CK, Smith GDW. MRS Bulletin 2009;34:725.
[24] Pereloma EV, Stohr RA, Miller MK, Ringer SP. Metall Mater Trans 2009;40A:3069.
[25] Sauvage X, Lefebvre W, Genevois C, Ohsaki S, Hono K. Scripta Mater 2009;60:1056.

[26] Al-Kassab T, Wollenberger H, Schmitz G, Kirchheim R, Tomography by Atom Probe in 'High Resolution Imaging and Specrtoscopy of Materials'. Eds. M. Rühle and F. Ernst, Springer Series in Materials Science, Vol. 50, Berlin Heidelberg: Springer-Verlag; 2000, pp 271-320
[27] Choi P, da Silva M, Klement U, Al-Kassab T, Kirchheim R. Acta Mater 2005; 53: 4473.
[28] Thermo-Calc Users' Guide, Version R, Thermo-Calc Software AB and Foundation of Computational Thermodynamics. Stockholm, Sweden;1995-2006.
[29] TCFE6 - TCS Steels/Fe-Alloys Database, Version 6.2.
[30] Borgenstam A, Engström A, Höglund L, Ågren J. J Phase Equilib 2000;21:269.
[31] Crusius S, Inden G, Knoop U, Höglund L, Ågren J. Z Metallk 1992;83:673.
[32] Franke P, Inden G. Z Metallk 1997;88:917.
[33] Bos C, Sietsma J. Scripta Mater 2007; 57:1085.
[34] Sietsma J, van der Zwaag S. Acta Mater 2004; 52: 4143.
[35] Calcagnotto M, Ponge D, Raabe D. Mater Sc Engin A 2010;527:2738.
[36] Ponge D, Millán J, Dmitrieva O, Sander B, Kostka A, Raabe D. Proc of the 2nd Int Symp on Steel Science (ISSS 2009) 2009, The Iron and Steel Inst of Japan:121

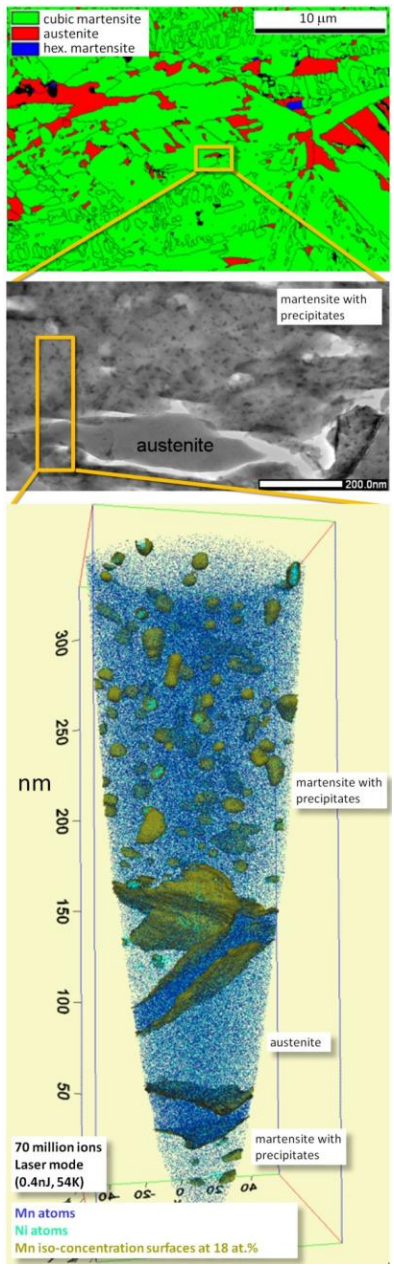

**Fig.1a**

Microstructure overview of the maraging-TRIP steel after quenching and subsequent aging (48 at 450 °C). The upper micrograph is an EBSD image where the cubic martensite is plotted green and the retained austenite red (the retained austenite was already present in the as-quenched state before aging). The middle image shows a TEM micrograph with precipitate-containing martensite and precipitate-free austenite. The bottom image shows an APT reproduction which includes both martensitic and austenitic zones. Ni atoms are given in cyan and Mn atoms in blue. The yellow iso-surfaces indicate 18 at.% Mn. Note that the three images properly reveal the hierarchy of the microstructure but the individual images were not exactly taken at the positions indicated.

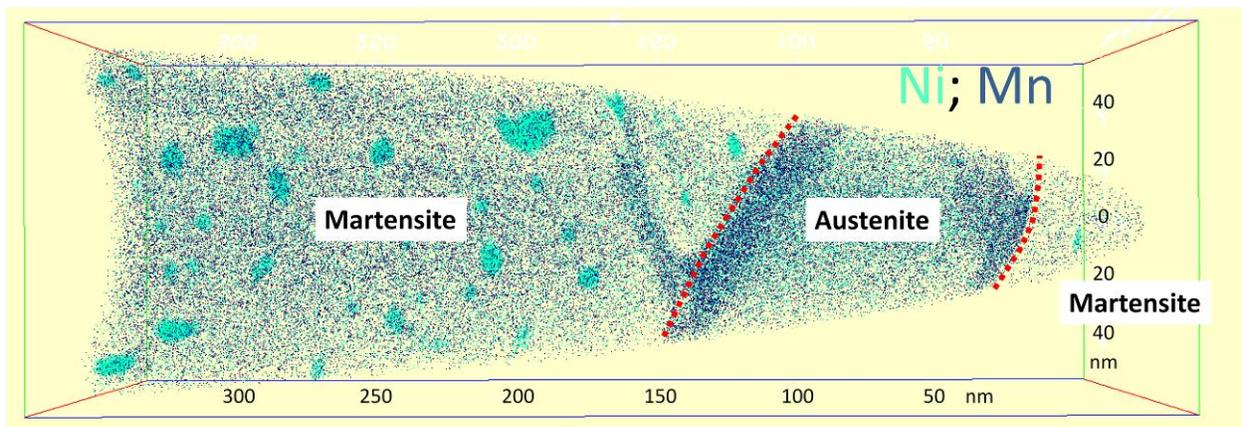

**Fig.1b** 20 nm thick middle layer slice through the APT reconstruction of the maraging TRIP steel shown in Fig. 1a. Ni atoms (cyan symbols) are accumulated in precipitates in the martensitic grains (left and right hand side). The precipitate-free austenite (right hand center) is bordered by plate-like zones that are characterized by strong Mn enrichment (blue symbols). Red dotted lines illustrate the suggested crystallographic positions of the phase boundaries between martensite and austenite.

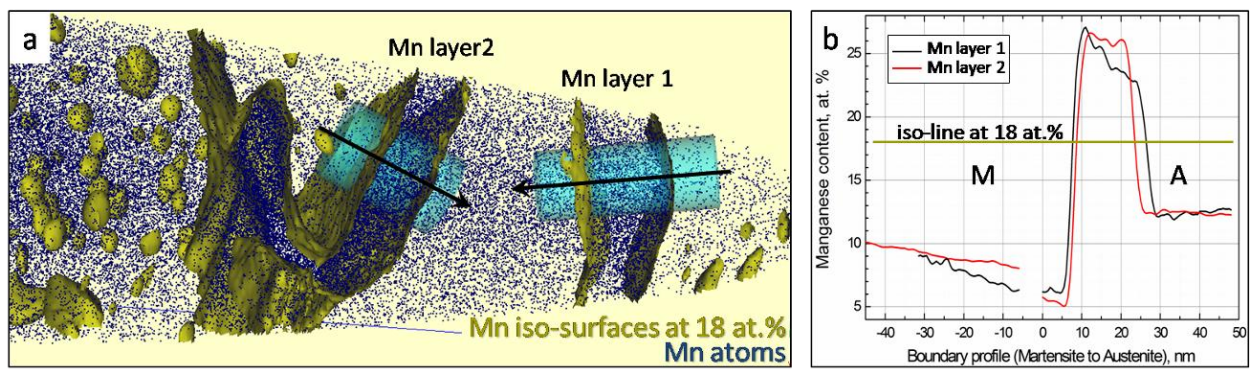

**Fig. 2:** Quantitative chemical analysis of the interface regions between martensite and austenite (APT results).

**2a**: Atomic map section showing both phase boundaries. Iso-concentration surfaces for the chemical distribution of Mn atoms (blue) were plotted at 18 at.% (yellow). One-dimensional profiles along the cylindrical units (cyan) provide chemical gradients of elements across the phase boundaries.

**2b**: Gradients in the Mn-content across the phase boundaries (martensite to austenite).

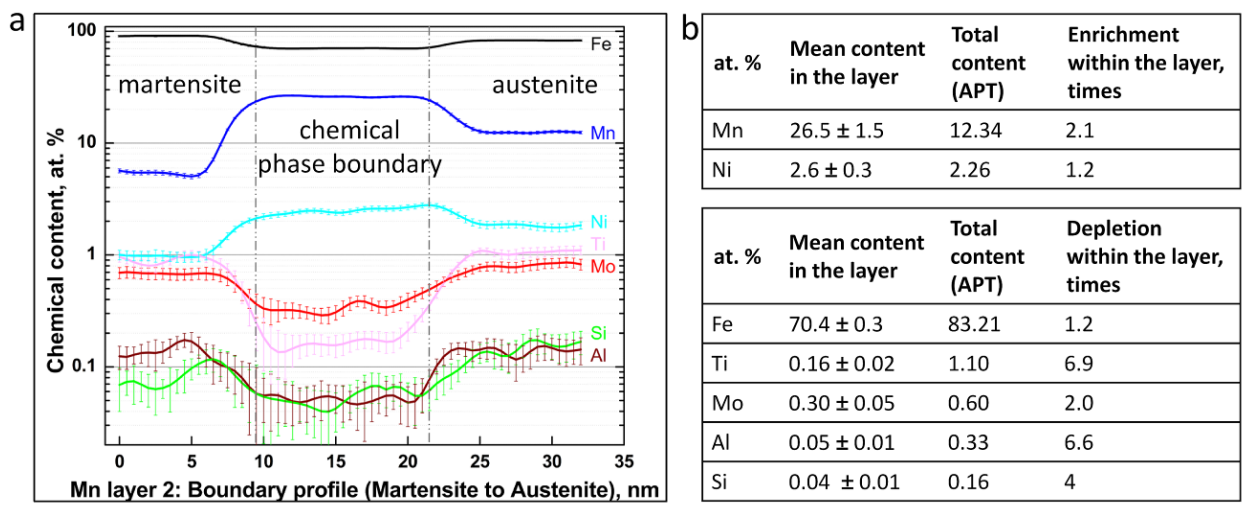

**Fig. 3** (Experimental APT results) **a:** Concentration profiles for all elements across one of the martensite/austenite phase boundaries (see interface referred to as 'Mn layer 2' in Fig. 2a). **b:** Quantitative characterization of the enrichment or depletion of the elements within the chemical phase boundary.

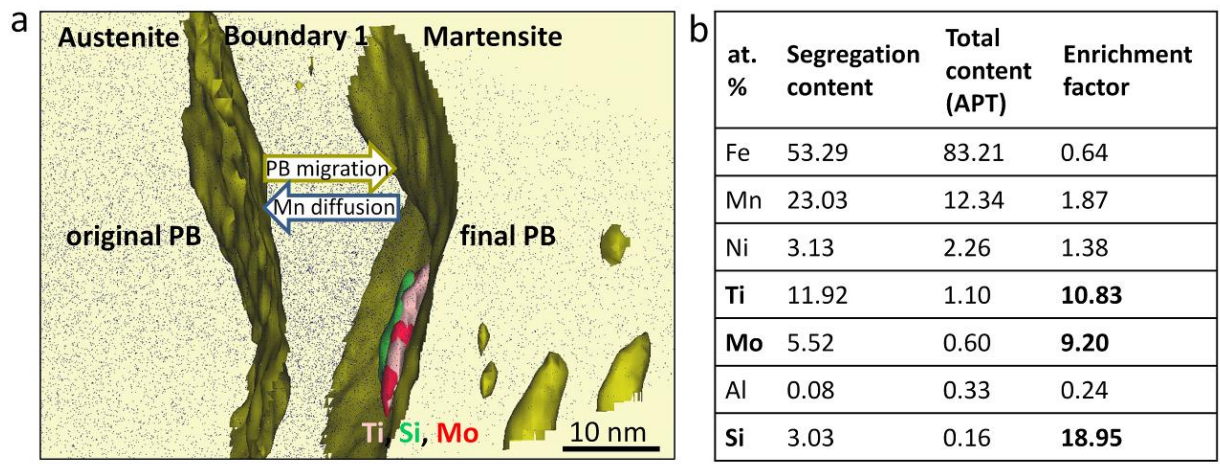

**Fig. 4:** Segregation within the austenite/martensite interface region (APT results).
**4a**: Atomic map section showing a phase boundary between austenite (left) and martensite (right). Iso-concentration surfaces plotted at 18 at.% Mn (dark yellow) correspond to the highest Mn gradient and indicates the positions of the original and the final phase boundaries (PB) (see text). Segregation at the final PB is revealed by plotting iso-concentration surfaces for Ti (at 8 at.%), Mo (at 5 at.%), and Si (at 3 at.%).
**4b**: Chemical composition of the Ti-Mo-Si-rich segregation estimated from the APT data.

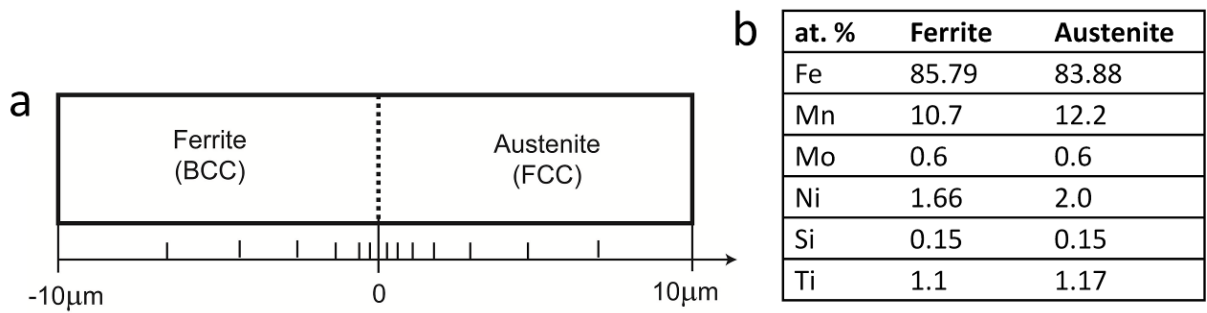

**Fig. 5  a:** Linear cell model set-up with ferrite and austenite as used in the DICTRA simulation. The spatial grid is defined in terms of a geometric series with a high density of grid points close to the interface.
**b:** Composition of the phases austenite and ferrite used as input for the DICTRA simulation.

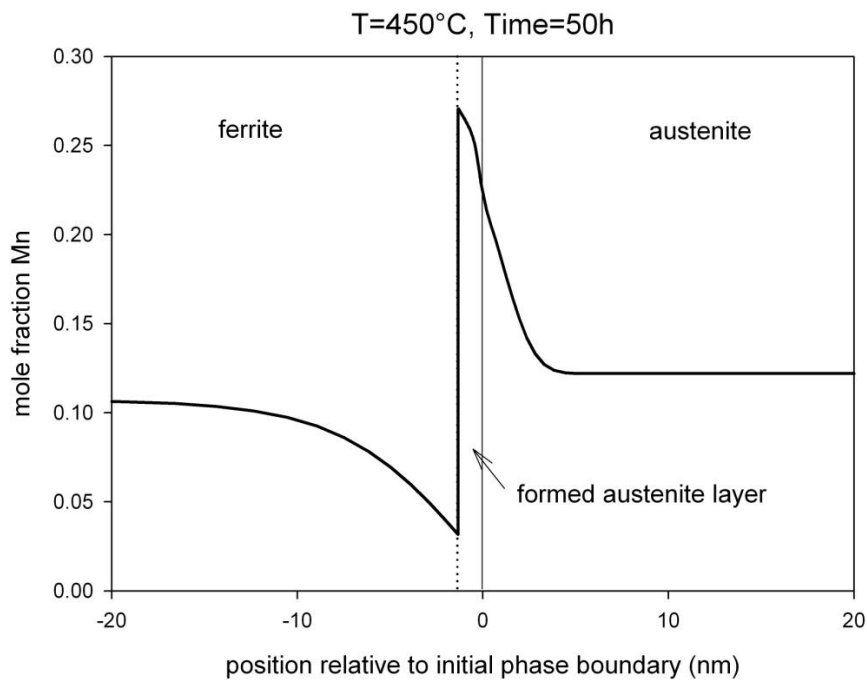

**Fig. 6:** DICTRA calculation of the Mn distribution at the martensite/austenite phase boundary. Martensite is thermodynamically and kinetically treated as ferrite. The calculation was done for 450°C (aging temperature). The result is shown for the time step 180000 sec (50 h).

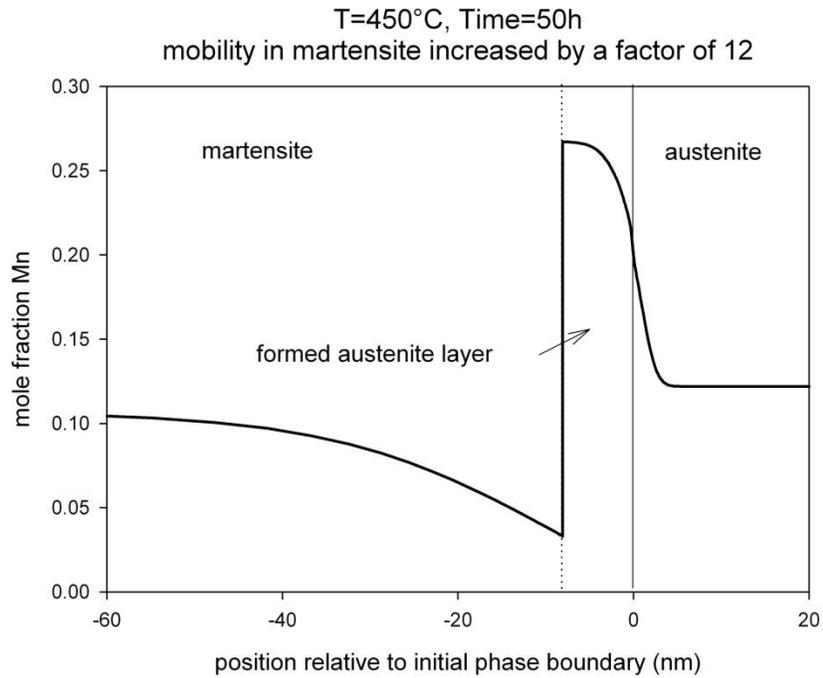

**Fig. 7a**

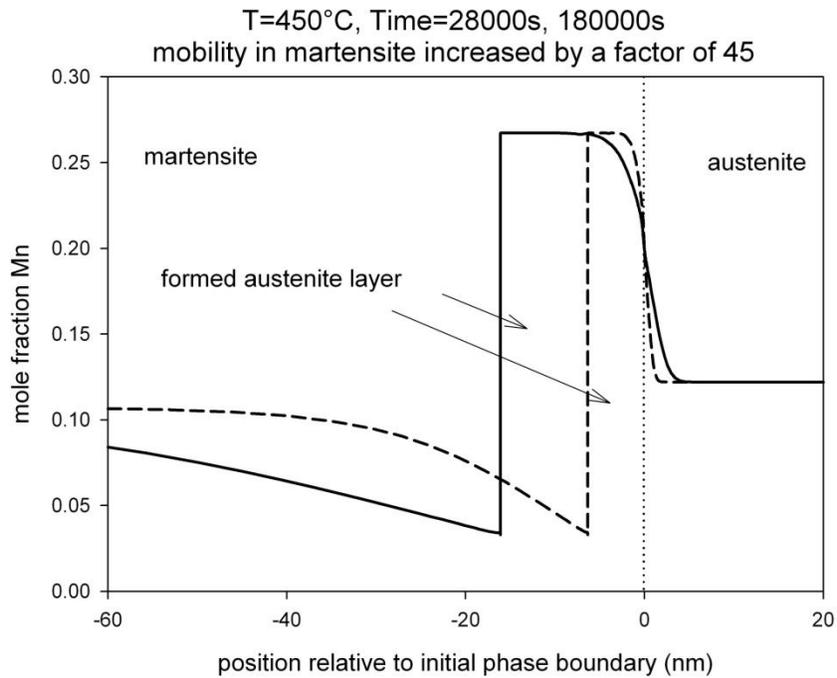

**Fig. 7b**

**Fig. 7:** DICTRA calculation of the Mn distribution at the martensite/austenite interface. Martensite is thermodynamically and kinetically treated as ferrite, but the mobility of the elements is increased by a factor of 12 (Fig. 7**a**) and 45 (Fig. 7**b**), respectively, owing to the high defect density in the martensite.

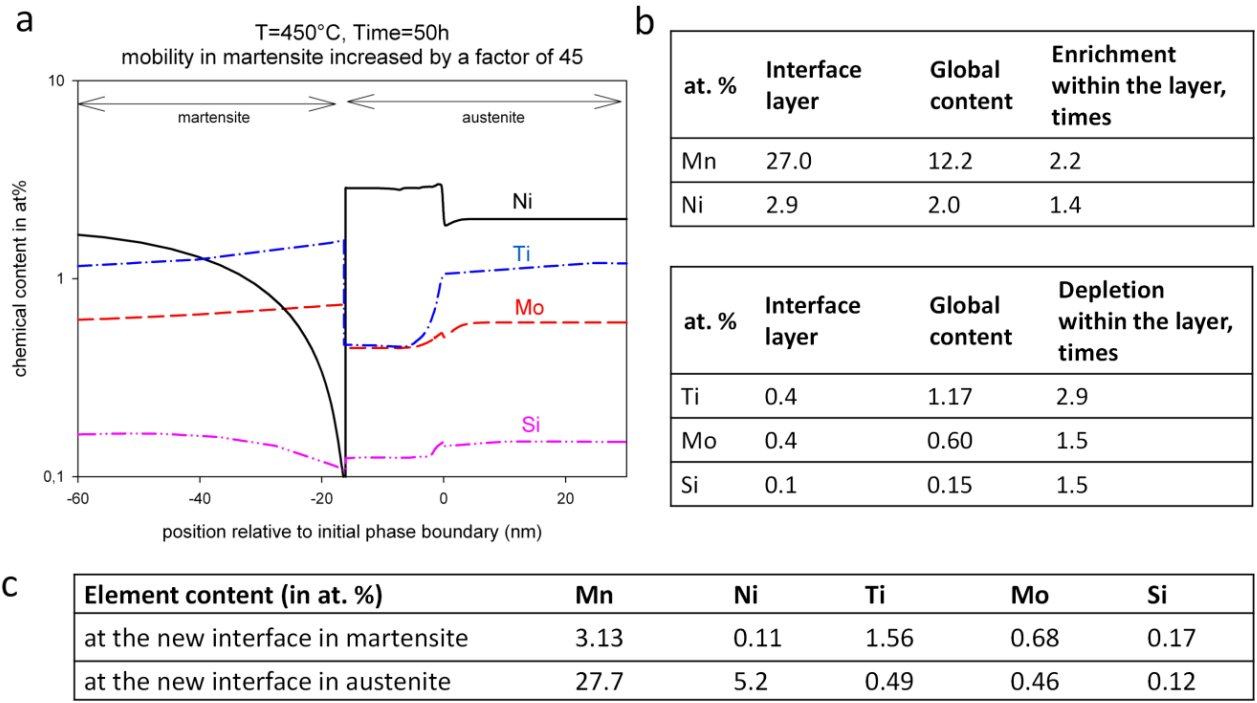

| Element content (in at. %) | Mn | Ni | Ti | Mo | Si |
|---|---|---|---|---|---|
| at the new interface in martensite | 3.13 | 0.11 | 1.56 | 0.68 | 0.17 |
| at the new interface in austenite | 27.7 | 5.2 | 0.49 | 0.46 | 0.12 |

**Fig. 8** (Results of DICTRA calculations) **a:** Composition profiles of all elements included in the DICTRA simulation. **b:** Quantitative characterization of the calculated profiles. **c:** Element contents at the new interface between martensite and austenite at 450°C for the given global composition.

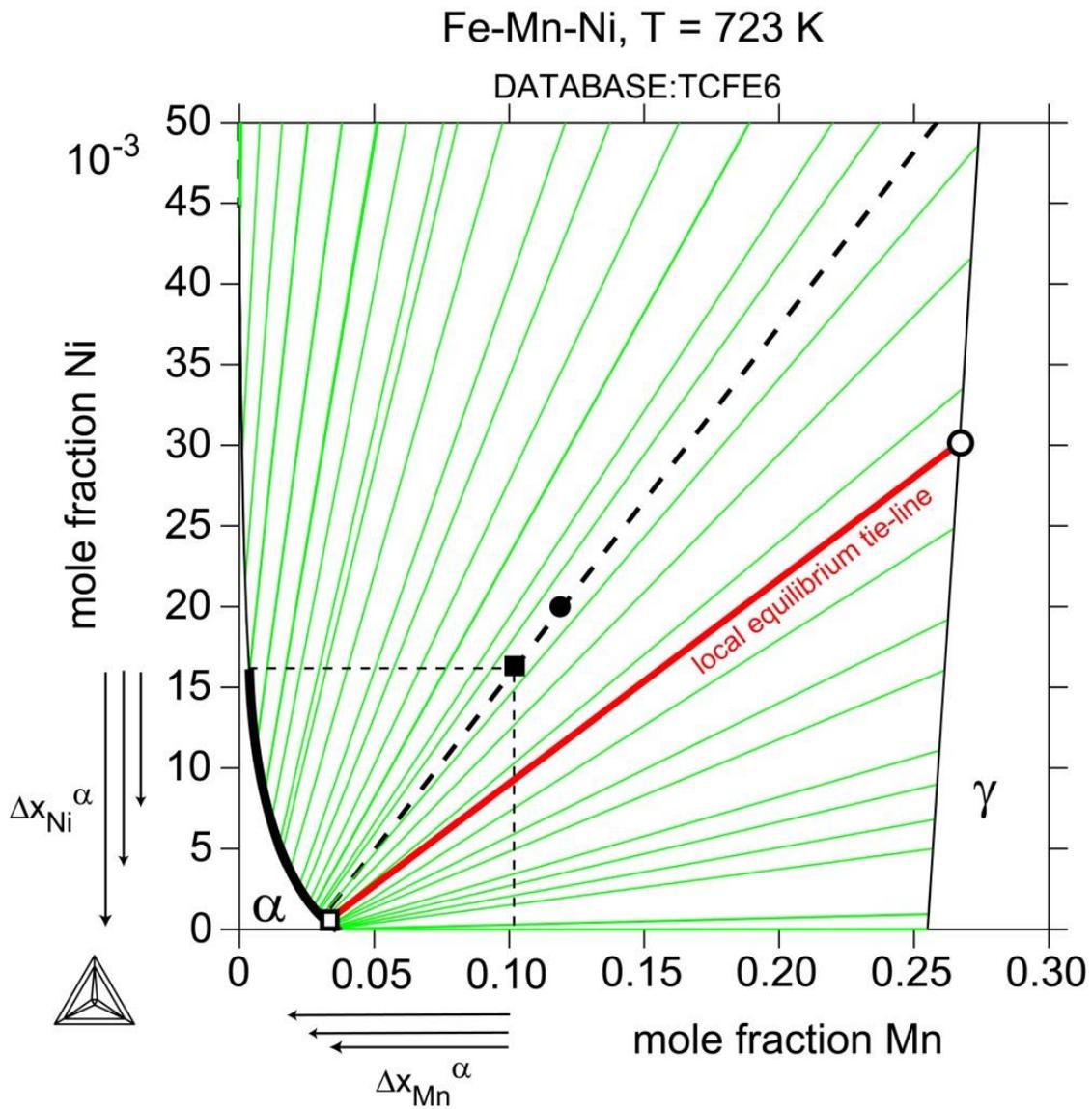

**Fig. 9:** Isothermal section of the Fe-Mn-Ni ternary system. The starting composition of austenite (filled circle) and martensite (filled square) are indicated. The global equilibrium tie-line is shown as broken line. The bold part of the ferrite phase boundary indicates the range of possible local equilibrium tie-lines.

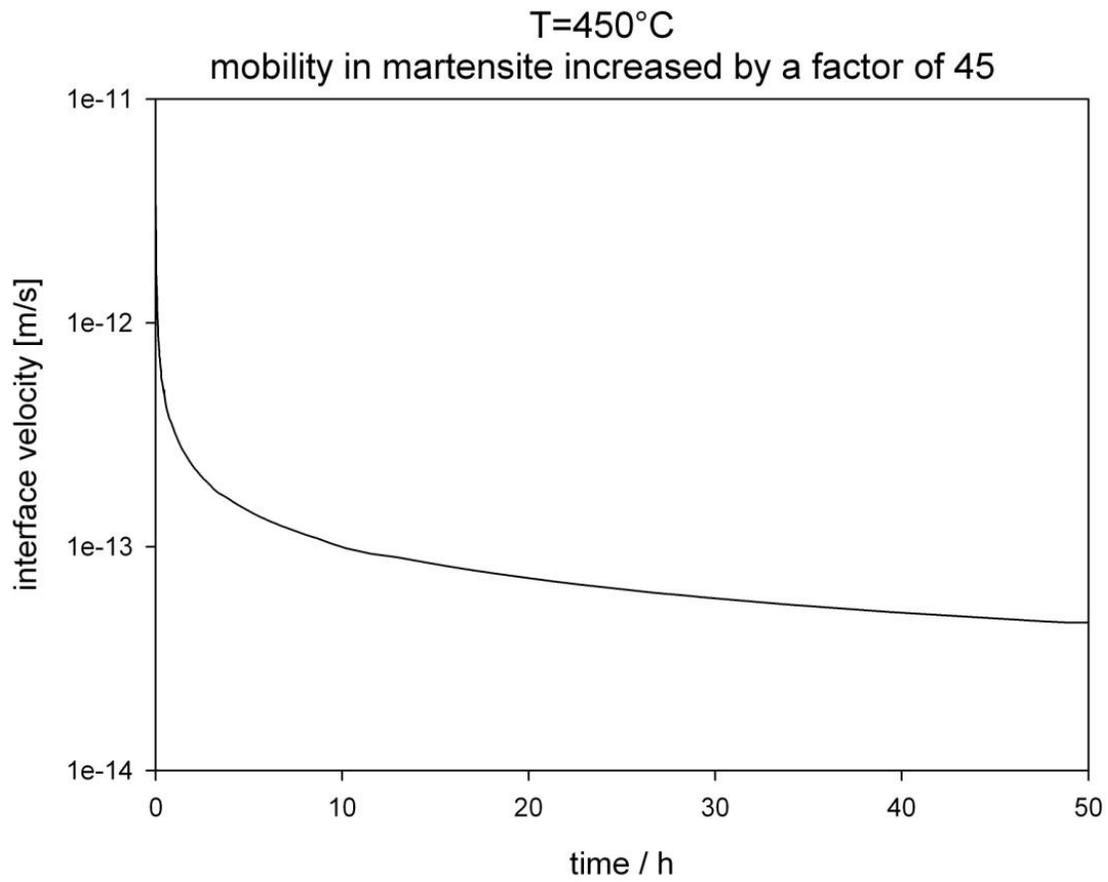

**Fig. 10:** Velocity of the α/γ interface as a function of time.

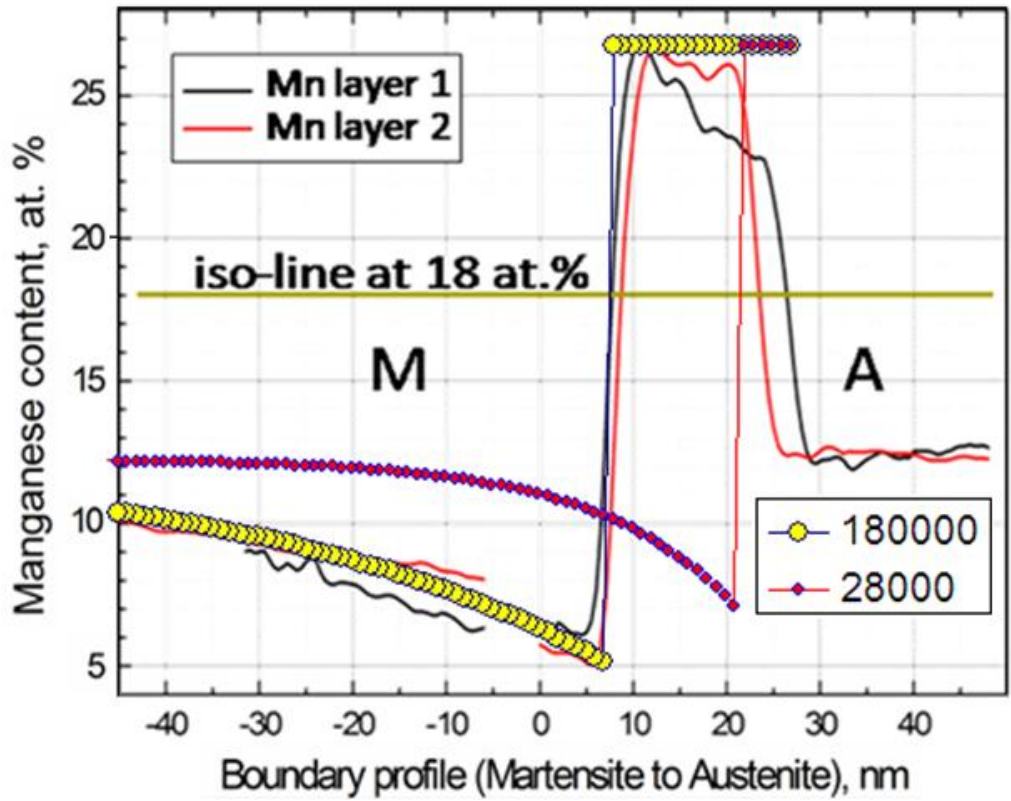

**Fig. 11** Results of the mixed-mode predictions of the Mn profile across the austenite/martensite interfaces. In contrast to the DICTRA simulation, here the interface mobility is taken into account [33]. The mixed-mode simulation results for two aging times (red points: 28000 s; yellow points: 180000 s) are plotted together with the experimental data (red and black lines, see Fig. 2b).

# Tables

| Chemical content, at. % | Total content (melt) | Total content (in APT) | Martensite | | | | Austenite |
|---|---|---|---|---|---|---|---|
| | | | total | matrix | particles | Enrichment factor | |
| Fe | 83.71 | 83.21 | 84.38 | 86.82 | 40.32 | 0.48 | 83.53 |
| **Mn** | **12.19** | **12.34** | **11.10** | 10.29 | 26.07 | 2.35 | **12.17** |
| Ni | 1.90 | 2.26 | 2.32 | 0.99 | 25.79 | **11.12** | 2.01 |
| Ti | 1.17 | 1.10 | 1.09 | 0.98 | 3.23 | 2.96 | 1.14 |
| Mo | 0.58 | 0.60 | 0.60 | 0.62 | 0.27 | 0.45 | 0.60 |
| Al | 0.31 | 0.33 | 0.34 | 0.14 | 4.08 | **12.0** | 0.38 |
| Si | 0.10 | 0.16 | 0.15 | 0.14 | 0.24 | 1.6 | 0.16 |
| C | 0.046 | 0.006 | 0.001 | 0.001 | 0 | - | 0.006 |

**Table 1** (Experimental results) Element composition of the alloy measured globally on the as-cast sample using wet chemical analysis (Total content melt) and obtained locally from the APT measurement on the specimen volume containing a martensite-austenite phase boundary of 450°C/48h aged steel (Total content APT; Martensite; Austenite). Enrichment factors are calculated as relation between the elemental content within the particles to the total content of element in the alloy.

| Phase | Mole fraction | Fe | Mn | Ni | Ti | Mo | Al | Si | C |
|---|---|---|---|---|---|---|---|---|---|
| BCC | 0.576 | 95.972 | 3.064 | 0.098 | 0.165 | 0.113 | 0.494 | 0.094 | - |
| FCC | 0.377 | 67.251 | 27.569 | 4.887 | 0.053 | 0.077 | 0.041 | 0.122 | - |
| Laves | 0.046 | 65.911 | 0.763 | - | 22.321 | 11.005 | - | - | - |
| TiC | 0.001 | - | - | - | 54.046 | - | - | - | 45.954 |

**Table 2** Equilibrium phases at 450°C in the investigated maraging-TRIP steel as obtained by ThermoCalc calculations quantified in terms of molar fractions (TCFE6 database).